# Pressure-enhanced spin-density-wave transition in double-layer nickelate $La_3Ni_2O_{7-\delta}$

Dan Zhao[1‡], Yanbing Zhou[1‡], Mengwu Huo[2‡], Yu Wang[1], Linpeng Nie[1], Ye Yang[1], Jianjun Ying[2,5], Meng Wang[2,*], Tao Wu[1,3,4,5,*] and Xianhui Chen[1,3,4,5,*]

1. Hefei National Research Center for Physical Sciences at the Microscale, University of Science and Technology of China, Hefei 230026, China

2. Center for Neutron Science and Technology, Guangdong Provincial Key Laboratory of Magnetoelectric Physics and Devices, School of Physics, Sun Yat-Sen University, Guangzhou 510275, China

3. CAS Key Laboratory of Strongly Coupled Quantum Matter Physics, Department of Physics, University of Science and Technology of China, Hefei 230026, China

4. Collaborative Innovation Center of Advanced Microstructures, Nanjing University, Nanjing 210093, China

5. Hefei National Laboratory, University of Science and Technology of China, Hefei 230088, China

‡These authors contributed equally to this work
*Correspondence to: wangmeng5@mail.sysu.edu.cn, wutao@ustc.edu.cn, chenxh@ustc.edu.cn

**Abstract:**

**Recently, a signature of high-temperature superconductivity above the liquid nitrogen temperature (77 K) was reported for $La_3Ni_2O_{7-\delta}$ under pressure. This finding immediately stimulated intense interest in the possible mechanism of high-$T_c$ superconductivity in double-layer nickelates. Notably, the pressure-dependent phase diagram inferred from transport measurements indicates that the superconductivity under high pressure emerges from the suppression of density-wave-like order at ambient pressure, which is similar to high-temperature superconductors. Therefore, clarifying the exact nature of the density-wave-like transition is important for determining the superconducting mechanism in double-layer nickelates. Here, nuclear magnetic resonance (NMR) spectroscopy of $^{139}$La nuclei was performed to study the density-wave-like transition in a single crystal of $La_3Ni_2O_{7-\delta}$. At high temperatures, two sets of sharp $^{139}$La NMR peaks are clearly distinguishable from a broad background signals, which are ascribed to La(1) sites from two bilayer Ruddlesden-Popper phases with different oxygen vacancy δ. As the temperature decreases, the temperature-dependent $^{139}$La NMR spectra and**

**nuclear spin-lattice relaxation rate ($1/T_1$) for both La(1) sites provide evidence of spin-density-wave (SDW) ordering below the transition temperature ($T_{SDW}$), which is approximately 150 K. The anisotropic splitting in the NMR spectra suggests the formation of a possible double spin stripe with magnetic moments aligned along the *c*-axis. Furthermore, we studied the pressure-dependent SDW transition up to ~ 2.7 GPa. Surprisingly, the $T_{SDW}$ inferred from NMR measurements of both La(1) sites increases with increasing pressure, which is opposite to the results from previous transport measurements under pressure and suggests an intriguing phase diagram between superconductivity and SDW. In contrast, the present $^{139}$La NMR is insensitive to the possible charge-density-wave (CDW) order in the Ni-O planes. All these results will be helpful for building a connection between superconductivity and magnetic interactions in double-layer nickelates.**

## 1. Introduction

The origin of high-temperature superconductivity remains a conundrum in condensed matter physics [1, 2]. The discovery of superconductivity up to 15 K in infinite-layer nickelate $NdNiO_2$ thin films has drawn extensive interest for the pursuit of high-$T_c$ superconducting mechanisms in nickelate materials [3]. Recent resonant inelastic X-ray scattering (RIXS) experiments have revealed significant antiferromagnetic interactions in superconducting $Nd_{1-x}Sr_xNiO_2$ thin films [4]. Moreover, muon spin rotation/relaxation (μSR) experiments have shown intrinsic magnetism in a series of superconducting infinite-layer nickelate thin films [5]. All these results are similar to those of cuprate superconductors [6-13], highlighting the key role of antiferromagnetic interactions in superconducting infinite-layer nickelates [14]. More recently, a signature of high-temperature superconductivity near 80 K was observed in the double-layer nickelate $La_3Ni_2O_{7-\delta}$ under high pressure [15]. Although characterization of the superconducting state remains limited due to the high-pressure environment, these findings have stimulated wide discussions on the electronic correlations and possible high-$T_c$ mechanism [16-31].

A density-wave-like transition at approximately 150 K was previously observed in $La_3Ni_2O_{7-\delta}$ at ambient pressure [32-37]. However, owing to the absence of detectable magnetic and charge ordering below 150 K in earlier experiments, the exact nature of this density-wave-like transition remains unclear [38]. In contrast, a similar density-wave-like transition has been observed in the trilayer nickelate $La_4Ni_3O_{10}$, in which an intertwined density wave involving both charge and spin order was experimentally revealed below the transition temperature (~ 136 K) [39]. Considering the similar band structures observed by angle-resolved photoemission spectroscopy (ARPES) [25, 40], understanding the different experimental results of density-wave-like transitions between $La_3Ni_2O_{7-\delta}$ and $La_4Ni_3O_{10}$ is not straightforward. More importantly, recent high-pressure transport measurements indicate that the density-wave-like transition in $La_3Ni_2O_{7-\delta}$ is continuously suppressed with increasing pressure, eventually disappearing as high-$T_c$ superconductivity emerges [15, 41, 42]. These results suggest an underlying correlation between superconductivity and density-wave-like order. Therefore, identifying the exact nature of the density-wave-like transition is a prerequisite for understanding the high-$T_c$ superconductivity in $La_3Ni_2O_{7-\delta}$ under high pressure. Nuclear magnetic resonance (NMR) is a sensitive local probe for detecting both spin and charge orders in high-$T_c$ superconductors [43, 44]. Here, we performed systematic NMR measurements of $^{139}$La nuclei in a single crystal of $La_3Ni_2O_{7-\delta}$. Our results provide clear evidence for a spin-density-wave (SDW) transition at ~ 150 K and reveal the pressure-dependent behavior of the SDW state.

## 2. Materials and methods

$La_3Ni_2O_7$ sample was fabricated by the high oxygen pressure floating zone technique and the details are described in [38]. A commercial NMR spectrometer from Thamway Co. Ltd was used for NMR and NQR measurements. An NMR-quality magnet from Oxford Instruments offers a uniform magnetic field. The external field at the sample's position was calibrated by $^{63}$Cu NMR of the NMR coil. All NMR spectra were measured by the standard spin echo method with a fast Fourier transform sum. The nuclear spin-lattice relaxation time $T_1$ of $^{139}$La nuclei was measured by the saturation method and the recovery of the nuclear magnetization $M(t)$ was fitted by a function with $M(t) = M_0 + M_1(0.028e^{-\left(\frac{t}{T_1}\right)^{\alpha}} + 0.178e^{-\left(\frac{6t}{T_1}\right)^{\alpha}} + 0.794e^{-\left(\frac{15t}{T_1}\right)^{\alpha}})$. The hydrostatic pressure is realized by using self-clamped BeCu piston-cylinder cell. The pressure transmitting medium is Daphne 7373 oil. The pressure was calibrated through the NQR frequency of $Cu_2O$ powder.

## 3. Site-selective $^{139}$La NMR of La(1) sites

As shown in Fig. 1b, $La_3Ni_2O_{7-\delta}$ at ambient pressure crystallizes into an orthorhombic phase, with a corner-connected $NiO_6$ octahedral layer separated by a La–O fluorite-type layer stacking along the *c* axis [36]. The La sites inside the corner-connected $NiO_6$ octahedral layer are defined as La(1), and the other La sites in the La-O fluorite-type layer are defined as La(2). Previous $^{139}$La NMR measurements of polycrystalline samples revealed distinct quadrupole frequencies ($\nu_Q$) for these two La sites: 2.2 MHz for La(1) and 5.6 MHz for La(2) [37, 45, 46], which is useful for obtaining site-selective $^{139}$La NMR spectra in single crystals. As shown in Fig. 1a, one piece of a $La_3Ni_2O_{7-\delta}$ single crystal with a size of ~ 2 mm × 1 mm was selected for $^{139}$La NMR measurement (see Sec. S1 in the Supplementary Materials for detailed sample information). Notably, although the intergrowth of $La_4Ni_3O_{10-\delta}$ or $La_2NiO_4$ is always involved in the sample growth of $La_3Ni_2O_{7-\delta}$ [47,48], the dominant phase in the present $La_3Ni_2O_{7-\delta}$ single crystal is the bilayer Ruddlesden-Popper (RP) phase (see Sec. S2 in the Supplementary Materials). The full NMR spectra were measured with magnetic fields ($B$) both perpendicular and parallel to the *ab* plane. The nuclear spin number for $^{139}$La nuclei is 7/2, and owing to nonzero quadrupole interactions, seven NMR transition lines are expected for the La(1) and La(2) sites. However, only one set of relatively sharp $^{139}$La NMR transition lines with $\nu_Q$ close to 2.2 MHz is resolved from the broad background, which is naturally assigned to La(1) sites. The absence of sharp NMR transition lines from the La(2) sites suggests that the present $La_3Ni_2O_{7-\delta}$ single crystal suffers from severe structural or ingredient inhomogeneity. In fact, recent multislice electron ptychography measurements indicate that the oxygen content in the $La_3Ni_2O_{7-\delta}$ single crystal is quite spatially inhomogeneous due to inner apical oxygen vacancies [49], which might be one reason for the low superconducting volume under pressure, as suggested by *ac* susceptibility measurements [50]. This inhomogeneity in oxygen content could significantly broaden the $^{139}$La NMR transition lines. Considering the predominant quadrupole broadening effect, the La(2) sites, with a larger quadrupole frequency ($\nu_Q \sim 5.6$ MHz), should have broader NMR transition lines than the La(1) sites ($\nu_Q \sim 2.2$ MHz). Furthermore, the principal axis of the electric field gradient (EFG) tensor for La(2) may not align with the $c$-axis, which could account for the absence of sharp NMR transition lines for La(2) sites. In the following, we focus only on the sharp NMR transition lines for the La(1) sites.

As shown in Fig. 1c, there is an additional split on each La(1) NMR transition line with $B//c$, which already exists above ~ 150 K. To clarify the origin of the unexpected splitting, we measured the field- and angle-dependent central transition lines for the La(1) sites, as shown in Fig. S3 (online). These results indicate that the splitting arises from two distinct La(1) sites with different EFG parameters and Knight shifts ($K$), which is not expected in a uniform orthorhombic structure (see Sec. S3 in the Supplementary Materials). If these two La(1) sites originate from a single phase, the corresponding crystal structure should contain two distinct La(1) chemical sites. However, all well-known structures of $La_3Ni_2O_7$ determined by XRD or neutron scattering (such as Amam, Fmmm, or Immm) contain only one La(1) chemical site. A natural explanation is that these two La(1) sites come from two different structural phases. Considering the inhomogeneous distribution of oxygen content in the present $La_3Ni_2O_{7-\delta}$ single crystal, as discussed above, these two La(1) sites might come from two bilayer RP phases with different oxygen contents. Recent neutron scattering experiments on $La_3Ni_2O_{7-\delta}$ polycrystals revealed that the inhomogeneous distribution of oxygen content in the bilayer RP phase can lead to two orthorhombic phases: Fmmm (No.69) and Amam (No.63) [51]. The Amam phase with an oxygen content of 6.79 accounts for 56.1% of the volume, and the Fmmm phase with an oxygen content of 6.90 accounts for 41.0% of the volume [51]. Density functional theory (DFT) calculations for $La_3Ni_2O_7$ with both Fmmm and Amam structures allowed us to calculate the EFG parameters for the La(1) sites in these structures (see Sec. S4 in the Supplementary Materials). Qualitatively, the DFT calculations give results for the quadrupole frequency and asymmetry factor that are consistent with the experimental observations (see Table S2 in the Supplementary Materials). Based on above DFT results, these two La(1) sites (labeled as La(1)a and La(1)b) are assigned to the bilayer RP phases with high oxygen content (Fmmm) and low oxygen content (Amam), respectively. Additionally, we have rechecked these two orthorhombic phases by measuring the nuclear quadrupole resonance (NQR) at the La(2) sites (see Sec. S5 in the Supplementary Materials), which further confirms the above conclusion drawn from the La(1) NMR results and provides an estimate of the total volume of these two orthorhombic phases (~ 68%). Furthermore, we noticed that previous $^{57}$Fe Mössbauer spectroscopy has also revealed two nonequivalent sites with population ratios close to 1:1 for $Fe^{3+}$ in $La_3Ni_{2-x}Fe_xO_{7\pm\delta}$ ($x$=0.05, 0.1) polycrystal samples [52], which might also be explained by these two orthorhombic phases. It should be emphasized that, regardless of how these two La(1) sites are attributed, the main conclusions of the SDW transition in the present work are almost not affected.

### 4. Evidence for an SDW transition at 150 K

Next, we systematically study the temperature-dependent evolution of central transition lines for both La(1)a and La(1)b sites. As shown in Fig. 2a, when the external magnetic field $B$ is applied perpendicular to the *ab* plane, both central transition lines for La(1)a and La(1)b significantly broaden below 150 K. The temperature-dependent linewidths for La(1)a and La(1)b are plotted in Fig. 2c, indicating a second-order phase transition at ~ 150 K. In contrast, when $B$ is applied parallel to the *ab* plane, the central transition lines for La(1)a and La(1)b overlap above 150 K and then exhibit a three-peak-like splitting below 150 K, as shown in Fig. 2b. If we define the splitting by the distance between the two farthest peaks, it is much larger than the broadening observed with *B*//*c*. The temperature-dependent splitting is plotted in Fig. 2d, which also supports a second-order phase transition at ~ 150 K. To further clarify the nature of the phase transition at ~ 150 K, we measured the field-dependent central transition lines with *B* parallel to the *ab* plane. As shown in Fig. 2e, above 150 K, the central transition line shows slight splitting as the magnetic field decreases from 15.5 to 7.5 T, which is ascribed to second-order quadrupole splitting for La(1)a and La(1)b. Since the temperature-dependent quadrupole frequencies for both La(1)a and La(1)b have no significant changes across 150 K (see Fig. S6 online), the large splitting below 150 K should be ascribed to significant magnetic splitting rather than to quadrupole splitting. Furthermore, we found that the large magnetic splitting below 150 K is almost field-independent, as shown in Fig. 2f, which further excludes Knight shift splitting and suggests the presence of a spontaneous internal field created by an SDW order. Similarly, we also checked the field-dependent central transition lines for *B* along different directions (see Sec. S7 in the Supplementary Materials). All these results demonstrate an SDW transition at ~ 150 K, which is also supported by recent μSR and RIXS experiments [53, 54]. In a previous study on the trilayer nickelate $La_4Ni_3O_{10}$, an intertwined density wave involving both spin and charge order was observed at ambient pressure [39]. Given the similar band structures of $La_3Ni_2O_7$ and $La_4Ni_3O_{10}$ [25, 40], a similar intertwined density wave order may also exist in $La_3Ni_2O_7$. Although NMR satellites can be utilized to detect charge ordering [44, 55], the temperature-dependent $^{139}$La NMR satellites do not show any signature for charge ordering at approximately 150 K (see Sec. S6 in the Supplementary Materials). If there is indeed any CDW order accompanied by the SDW order in $La_3Ni_2O_7$, it is beyond the sensitivity of the present $^{139}$La NMR measurements on La(1) sites. Below, we consider only pure SDW models to explain our present NMR results.

By careful in-plane rotation, two orthorhombic domains were identified by the present NMR measurements (see Sec. S8 in the Supplementary Materials). All these NMR spectra obtained with magnetic fields along different directions (Fig. 2 and Fig. S11 online) can be used to constrain the possible magnetic structure of the SDW state below 150 K. As shown in Fig. 2g and 2h, by considering a pseudo-tetragonal unit cell with a lattice constant of ~ 3.83 Å, a double spin stripe model with a wavevector $q = (\pi/2, \pi/2)$ is proposed to account for the present NMR results. This magnetic model is consistent with the results of recent RIXS and neutron scattering experiments [51, 54]. Here, we consider an antiferromagnetic coupling between the neighboring Ni-O planes in each bilayer, with the spin polarization occurring along the out-of-plane direction. According to this model, as shown in Fig. 2i–l, the transferred hyperfine fields from the nearest neighboring Ni sites to the La(1)a and La(1)b sites split the $^{139}$La NMR spectra only when $B//ab$, and not when $B\perp ab$, which explains the observed $^{139}$La NMR spectra shown in Fig. 2 and Fig. S15 (online). More discussion of the possible magnetic structure is provided in the supplementary materials (see Sec. S9 in the Supplementary Materials). Notably, the transferred hyperfine fields for La(1)a and La(1)b are quite different, as reflected in the above analysis of the NMR spectra. Based on the analysis of the hyperfine fields, we can estimate the static magnetic moments at each Ni site to be approximately 0.08 Bohr magneton ($\mu_B$) for Ni around La(1)a and 0.018 $\mu_B$ for Ni around La(1)b (see Sec. S10 in the Supplementary Materials). Considering the abovementioned bilayer RP phases with different oxygen contents, this result suggests that the inner apical oxygen is helpful for stabilizing static magnetic moments in the SDW state. This is consistent with a recent density function theory calculation [56], which shows that the magnitude of the magnetic moment is strongly affected by the absence of inner apical oxygens in $La_3Ni_2O_7$.

## 5. Pressure-dependent SDW transition

We also measured the temperature-dependent nuclear spin-lattice relaxation rate ($1/T_1$) which further confirms an SDW transition at ~ 150 K. As shown in Fig. 3a, the temperature-dependent $1/T_1T$ measurements of La(1)a and La(1)b exhibit second-order transition behavior at ~ 150 K, which is roughly consistent with previous NMR investigations of polycrystalline samples [37,46]. Interestingly, the present measurement on single crystal reveals different temperature-dependent behaviors for La(1)a and La(1)b above 150 K. For La(1)a, the $1/T_1T$ is significantly enhanced below 200 K and then reaches a maximum at ~ 150 K, which is different from the roughly $T$-linear behavior observed in the Knight shift (see Sec. S11 in the Supplementary Materials). This behavior in $1/T_1T$ indicates significant spin fluctuations above 150 K, which is consistent with an SDW transition and is widely observed in iron-based superconductors [43]. In contrast, the $1/T_1T$ for La(1)b is continuously suppressed with decreasing temperature towards 150 K and then exhibits a weak peak-like behavior within a few Kelvins at ~ 150 K. The overall temperature-dependent $1/T_1T$ for La(1)b is similar to the previous La(1) NMR results for polycrystalline samples [37, 46]. How can the different $1/T_1T$ behaviors for La(1)a and La(1)b above 150 K be understood? As we discussed before, the static magnetic moments experienced by La(1)a (~ 0.08 $\mu_B$) are much larger than those experienced by La(1)b (~ 0.018 $\mu_B$). If we consider that the observed spin fluctuations above 150 K are related to the static magnetic moments below 150 K, the contribution of spin fluctuations to $1/T_1T$ in La(1)a should be roughly 20 times larger than that in La(1)b. Therefore, the significant spin fluctuations experienced by La(1)a are related to a larger magnetic moment below 150 K, which is strongly dependent on the oxygen content. Below the SDW transition temperature ($T_{\mathrm{SDW}} \sim 150$ K), the $1/T_1T$ for both La(1)a and La(1)b initially decreases rapidly and then continues to decrease as the temperature approaches 50 K. Moreover, the stretched exponent, $r$, extracted from fitting the $T_1$ decay and indicative of $T_1$ inhomogeneity, decreases below $T_{SDW}$ (Fig. 3b). This indicates inhomogeneous spin dynamics in the SDW state, which might be related to the abovementioned inhomogeneous distribution of the oxygen content. Below 50 K, additional spin dynamics appear in the temperature-dependent $1/T_1T$ and lead to field-dependent peak behavior at lower temperatures (Fig. 3c). This field-dependent behavior can be ascribed to the dynamic slowing down of fluctuating magnetic moments, which is usually observed in glassy magnetism [8, 12, 57]. A similar phenomenon has been observed in slightly doped $La_2CuO_4$, where the spin freezing behavior induced by hole doping is superimposed on the characteristic response for long-range antiferromagnetic order [58]. In our case, the apical oxygen vacancies might play a role similar to that of hole doping in $La_2CuO_4$, which locally disturbs the long-range SDW and generates spin freezing behavior at low temperatures. In fact, evidence for long-range SDW order is still lacking in recent neutron scattering experiments [51]. The residual spin dynamics observed by nuclear spin-lattice relaxation at low temperatures might be consistent with the absence of long-range SDW order.

Next, we further measured the pressure dependence of the SDW transition in $1/T_1T$. As shown in Fig. 4a and 4b, the temperature-dependent $1/T_1T$ was measured under different hydrostatic pressures up to 2.7 GPa (see details in Sec. S12 of the Supplementary Materials). Surprisingly, the $T_{\mathrm{SDW}}$ determined by the peak temperature in $1/T_1T$ does not decrease but instead slightly increases with increasing pressure. Previous high-pressure transport measurements of $La_3Ni_2O_{7-\delta}$ single crystal samples from the same batch indicate that the density-wave-like transition, determined by the resistivity anomaly, is continuously suppressed with increasing pressure and disappears above 2.0 GPa, as shown in Fig. 4c. This discrepancy suggests that the density-wave-like transition in transport is likely due to a different type of density-wave order rather than the SDW order. In addition to the SDW order, the CDW order was also proposed in previous studies [38, 60]. Recent optical conductivity measurements revealed that an energy gap opens below 115 K in the $La_3Ni_2O_{7-\delta}$ single crystal, suggesting the formation of a CDW order [29]. It would be very interesting for future optical conductivity experiments to verify whether the energy gap would be closed or not above 2.0 GPa. Furthermore, our present results also indicate that the pressure-dependent SDW transition has no significant effect on the charge transport, which is quite consistent with recent ARPES experiments [25]. This finding is inconsistent with the idea of a Fermi-surface-nesting driven SDW transition. If the resistivity anomaly is indeed ascribed to the development of the CDW order, then the SDW and CDW orders exhibit distinct pressure-dependent evolution in $La_3Ni_2O_{7-\delta}$ as shown in Fig. 4c. Considering that the SDW transition temperature is higher than the CDW transition temperature, the CDW order might be driven by the SDW order, which has already been observed in conventional SDW systems [61]. As we discussed before, the significant number of oxygen vacancies in $La_3Ni_2O_{7-\delta}$ can affect the SDW order. The distinct pressure-dependent evolution of the SDW and CDW orders might also be related to the oxygen vacancy, which deserves further investigation for $La_3Ni_2O_{7-\delta}$ with reduced oxygen vacancies.

## 6. Discussion and conclusion

Finally, the robust SDW order against pressure suggests an intriguing picture for the appearance of superconductivity under high pressure. The superconducting phase with the maximum $T_c$ might appear close to the SDW phase, as shown in Fig. 4c. This phase diagram is distinct from that of high-$T_c$ cuprate and iron-based superconductors [1, 2, 62], but highly similar to the pressure-dependent phase diagram of the $Cs_3C_{60}$ superconductor [63]. Considering that there is no sudden change in structure with increasing pressure, the above picture suggests a strong connection between superconductivity and SDW order. In unconventional superconductors, magnetic exchange interactions are widely regarded as crucial for superconducting pairing [1, 2]. Our results suggest that the superconductivity in $La_3Ni_2O_7$ is potentially linked to magnetism. However, further experiments under higher pressure are needed to confirm whether superconductivity arises from the gradual suppression of long-range antiferromagnetic order, as observed in other unconventional superconductors, such as cuprate and iron-based superconductors.

*Notice:* When our manuscript was initially submitted, a pressure-dependent μSR study also appeared in arXiv [64], which is consistent with our present NMR results on the pressure-dependent SDW transition.

## Conflict of interest

The authors declare that they have no conflict of interest.

## Acknowledgments

We thank the valuable discussion with Kun Jiang. This work was supported by the National Key R&D Program of the MOST of China (2022YFA1602601, 2023YFA1406500), the National Natural Science Foundation of China (12494592, 12034004, 12161160316, 12325403, 12174454, 12204450, and 12488201), the Strategic Priority Research Program of Chinese Academy of Sciences (XDB25000000), the Chinese Academy of Sciences under contract No. JZHKYPT-2021-08, the CAS Project for Young Scientists in Basic Research (YBR-048), the Innovation Program for Quantum Science and Technology (2021ZD0302800), the Guangdong Basic and Applied Basic Research Funds (2021B1515120015), the Guangzhou Basic and Applied Basic Research Funds (2024A04J6417), the Guangdong Provincial Key Laboratory of Magnetoelectric Physics and Devices (2022B1212010008), and the Fundamental Research Funds for the Central Universities (WK9990000110).

## Author contributions

Tao Wu and Xianhui Chen conceived the project. Dan Zhao, Yanbing Zhou, Yu Wang, and Linpeng Nie performed NMR experiments. Jianjun Ying performed resistivity measurements. Ye Yang performed first-principles calculations. Mengwu Huo and Meng Wang grew the single crystals. Dan Zhao and Tao Wu interpreted the results. Dan Zhao and Tao Wu wrote the manuscript with input from all the authors. All authors discussed the results and commented on the manuscript.

## Appendix A. Supplementary material

Supplementary data to this article can be found online at https://doi.org/10.1016/j.scib.2025.02.019

## Reference

[1] Lee PA, Nagaosa N, Wen XG. Doping a Mott insulator: Physics of high-temperature superconductivity. Rev Mod Phys 2006; 78: 17-85.
[2] Keimer, B, Kivelson, SA., Norman, M. R, et al. From quantum matter to high-temperature superconductivity in copper oxides. Nature 2015; 518: 179-186.

[3] Li DF, Lee K, Wang BY, et al. Superconductivity in an infinite-layer nickelate. Nature 2019; 572: 624-627.
[4] Lu H, Rossi M, Nag A, et al. Magnetic excitations in infinite-layer nickelates. Science 2021; 373: 213-216.
[5] Fowlie J, Hadjimichael M, Martins MM, et al. Intrinsic magnetism in superconducting infinite-layer nickelates. Nat Phys 2022; 18: 1043-1047.
[6] Julien M.-H. Magnetic order and superconductivity in $La_{2-x}Sr_xCuO_4$: a review. Physica B Condens Matter 2003; 329: 693-696.
[7] Niedermayer Ch, Bernhard C, Blasius T, et al. Common phase diagram for antiferromagnetism in $La_{2-x}Sr_xCuO_4$ and $Y_{1-x}Ca_xBa_2Cu_3O_6$ as seen by muon spin rotation. Phys Rev Lett 1998; 80: 3843.
[8] Julien MH, Borsa F, Carretta P, et al. Charge segregation, cluster spin glass, and superconductivity in $La_{1.94}Sr_{0.06}CuO_4$. Phys Rev Lett 1999; 83: 604.
[9] Panagopoulos C, Tallon JL, Rainford BD, et al. Low-frequency spins and the ground state in high-$T_c$ cuprates. Solid State Commun 2003; 126: 47-55.
[10] Haug D, Hinkov V, Sidis Y, et al. Neutron scattering study of the magnetic phase diagram of underdoped $YBa_2Cu_3O_{6+x}$. New J Phys 2010; 12: 105006.
[11] Risdiana, Adachi T, Oki N, et al. Cu spin dynamics in the overdoped regime of $La_{2-x}Sr_xCu_{1-y}Zn_yO_4$ probed by muon spin rotation. Phys Rev B 2008; 77: 054516. .
[12] Wu T, Mayaffre H, Krämer S, et al. Magnetic-field-enhanced spin freezing on the verge of charge ordering in $YBa_2Cu_3O_{6.45}$, Phys Rev B 2013; 88: 014511.
[13] Frachet M, Vinograd I, Zhou R, et al. Hidden magnetism at the pseudogap critical point of a cuprate superconductor. Nat Phys 2020; 16: 1064-1068.
[14] Zhao D, Zhou YB, Fu Y, et al. Intrinsic spin susceptibility and pseudogaplike behavior in infinite-layer $LaNiO_2$. Phys Rev Lett 2021; 126: 197001.
[15] Sun H, Huo M, Hu X, et al. Signatures of superconductivity near 80 K in a nickelate under high pressure. Nature 2023; 621: 493-498.
[16] Gu Y, Le C, Yang Z, et al. Effective model and pairing tendency in bilayer Ni-based superconductor $La_3Ni_2O_7$. arXiv: 2306.07275; 2023.
[17] Sakakibara H, Kitamine N, Ochi M, et al. Possible high $T_c$ superconductivity in $La_3Ni_2O_7$ under high pressure through manifestation of a nearly half-filled bilayer hubbard model. Phys Rev Lett 2024; 132: 106002.
[18] Yang YF, Zhang GM, Zhang FC. Interlayer valence bonds and two-component theory for high-Tc superconductivity of $La_3Ni_2O_7$ under pressure. Phys Rev B 2023; 108: L201108.
[19] Qin Q, Yang YF. High-$T_c$ superconductivity by mobilizing local spin singlets and possible route to higher Tc in pressurized $La_3Ni_2O_7$, Phys Rev B 2023; 108: L140504.
[20] Shen Y, Qin MP, Zhang GM. Effective bi-layer model hamiltonian and density-matrix renormalization group study for the high-$T_c$ superconductivity in $La_3Ni_2O_7$ under high pressure. Chin Phys Lett 2023; 40: 127401.
[21] Yang QG, Wang D, Wang QH. Possible $s^{\pm}$-wave superconductivity in $La_3Ni_2O_7$. Phys Rev B 2023; 108: L140505.
[22] Qu XZ, Qu DW, Chen JL, et al. Bilayer $t-J-J_{\perp}$ model and magnetically mediated pairing in the pressurized nickelate $La_3Ni_2O_7$. Phys Rev Lett 2024; 132: 036502.
[23] Zhang Y, Lin LF, Moreo A, et al. Electronic structure, dimer physics, orbital-selective behavior, and magnetic tendencies in the bilayer nickelate superconductor $La_3Ni_2O_7$ under pressure, Phys Rev B 2023; 108: L180510.

[24] Liu YB, Mei JW, Ye F, et al. $s^{\pm}$-Wave Pairing and the destructive role of apical-oxygen deficiencies in $La_3Ni_2O_7$ under pressure. Phys Rev Lett 2023; 131: 236002.
[25] Yang J, Sun H, Hu X, et al. Orbital-dependent electron correlation in double-layer nickelate $La_3Ni_2O_7$. Nat Commun 2024; 15: 4373.
[26] Lu C, Pan Z, Yang F, et al. Interplay of two $E_g$ orbitals in superconducting $La_3Ni_2O_7$ under pressure. Phys Rev B 2024; 110: 094509.
[27] Luo ZH, Hu X, Wang M, et al. Bilayer two-orbital model of $La_3Ni_2O_7$ under pressure. Phys Rev Lett 2023; 131: 126001.
[28] Tian YH, Chen Y, Wang JM, et al. Correlation effects and concomitant two-orbital $s\pm$-wave superconductivity in $La_3Ni_2O_7$ under high pressure. Phys Rev B 2024; 109: 165154.
[29] Liu Z, Huo M, Li J, et al. Electronic correlations and partial gap in the bilayer nickelate $La_3Ni_2O_7$. Nat Commun 2024; 15: 7570.
[30] Christiansson V, Petocchi F, Werner P. Correlated electronic structure of $La_3Ni_2O_7$ under pressure. Phys Rev Lett 2023; 131: 206501.
[31] Chen JL, Yang F, Li W, et al. Orbital-selective superconductivity in the pressurized bilayer nickelate $La_3Ni_2O_7$: an infinite projected entangled-pair state study. Phys Rev B 2024; 110: L04111.
[32] Sreedhar K, McElfresh M, Perry D, et al. Low-temperature electronic properties of the $La_{n+1}Ni_nO_{3n+1}$ ($n$ = 2, 3, and $\infty$) system: evidence for a crossover from fluctuating-valence to fermi-liquid-like behavior. J Solid State Chem 1994; 110: 208-215.
[33] Zhang Z, Greenblatt M, Goodenough J. Synthesis, structure, and properties of the layered perovskite $La_3Ni_2O_{7-\delta}$. J Solid State Chem 1994; 108: 402-409.
[34] Taniguchi S, Nishikawa T, Yasui Y, et al. Transport, magnetic and thermal properties of $La_3Ni_2O_{7-\delta}$. J Phys Soc Jpn 1995; 64: 1644-1650.
[35] Kobayashi Y, Taniguchi S, Kasai M, et al. Transport and magnetic properties of $La_3Ni_2O_{7-\delta}$ and $La_4Ni_3O_{10-\delta}$. J Phys Soc Jpn 1996; 65: 3978-3982.
[36] Ling CD, Argyriou DN, Wu G, et al. Neutron diffraction study of $La_3Ni_2O_7$: structural relationships among $n$=1, 2, and 3 phases $La_{n+1}Ni_nO_{3n+1}$. J Solid State Chem 2000; 152: 517-525.
[37] Fukamachi T, Kobayashi Y, Miyashita T, et al. $^{139}$La NMR studies of layered perovskite systems $La_3Ni_2O_{7-\delta}$ and $La_4Ni_3O_{10}$. J Phys Chem Solids 2001; 62: 195-198.
[38] Liu Z, Sun H, Huo M, et al. Evidence for charge and spin density waves in single crystals of $La_3Ni_2O_7$ and $La_3Ni_2O_6$. Sci China Phys Mech Astron 2023; 66: 217411.
[39] Zhang J, Phelan D, Botana AS, et al. Intertwined density waves in a metallic nickelate. Nat Commun 2020; 11: 6003.
[40] Li HX, Zhou XQ, Nummy T, et al. Fermiology and electron dynamics of trilayer nickelate $La_4Ni_3O_{10}$. Nat Commun 2017; 8: 704.
[41] Zhang Y, Su D, Huang Y, et al. High-temperature superconductivity with zero resistance and strange-metal behavior in $La_3Ni_2O_{7-\delta}$, Nat Phys 2024; 20: 1269-1273.
[42] Wang G, Wang NN, Shen XL, et al. Pressure-induced superconductivity in polycrystalline $La_3Ni_2O_{7-\delta}$, Phys Rev X 2024; 14: 011040.
[43] Kitagawa K, Katayama N, Ohgushi K, et al. Commensurate itinerant antiferromagnetism in $BaFe_2As_2$: $^{75}$As-NMR studies on a self-flux grown single crystal. J Phys Soc Jpn 2008; 77: 114709.
[44] Wu T, Mayaffre H, Krämer S, et al. Magnetic-field-induced charge-stripe order in the high-temperature superconductor $YBa_2Cu_3O_y$. Nature 2011; 477: 191-194.
[45] Fukamachi T, Oda K, Kobayashi Y, et al. Studies on successive electronic state changes in systems with $NiO_2$ planes –$^{139}$La-NMR/NQR. J Phys Soc Jpn 2001; 70: 2757.

[46] Kakoi M, Oi T, Ohshita Y, et al. Multiband metallic ground state in multilayered nickelates $La_3Ni_2O_7$ and $La_4Ni_3O_{10}$ revealed by $^{139}$La −NMR at ambient pressure. J Phys Soc Jpn 2024; 93: 053702.
[47] Puphal P, Reiss P, Enderlein N, et al. Unconventional crystal structure of the high-pressure superconductor $La_3Ni_2O_7$. Phys Rev Lett 2024; 133: 146002.
[48] Chen X, Zhang J, Thind AS, et al. Polymorphism in ruddlesden–popper $La_3Ni_2O_7$: discovery of a hidden phase with distinctive layer stacking. J Am Chem Soc 2024; 146: 3640-3645.
[49] Dong Z, Huo M, Li J, et al. Visualization of oxygen vacancies and self-doped ligand holes in $La_3Ni_2O_{7-\delta}$. Nature 2024; 630: 847-852.
[50] Zhou Y, Guo J, Cai S, et al. Investigations of key issues on the reproducibility of high-$T_c$ superconductivity emerging from compressed $La_3Ni_2O_7$. Matter Radiat Extremes 2025; 10: 027801.
[51] Xie T, Huo M, Ni X, et al. Strong interlayer magnetic exchange coupling in $La_3Ni_2O_{7-\delta}$ revealed by inelastic neutron scattering, Sci Bull 2024; 69: 3221-3227.
[52] Kiseleva EA, Gaczynski P, Eckold G, et al. Investigations into the structure of $La_3Ni_{2-x}Fe_xO_{7\pm\delta}$. Chimica Techno Acta 2019; 6: 51-57.
[53] Chen KW, Liu XQ, Jiao JC, et al. Evidence of spin density waves in $La_3Ni_2O_{7-\delta}$. Phys Rev Lett 2024; 132: 256503.
[54] Chen X, Choi J, Jiang Z, et al. Electronic and magnetic excitations in $La_3Ni_2O_7$. Nat Commun 2024; 15: 9597.
[55] Kawasaki S, Li Z, Kitahashi M, et al. Charge-density-wave order takes over antiferromagnetism in $Bi_2Sr_{2-x}La_xCuO_6$ superconductors, Nat Commun 2017; 8: 1267.
[56] Wang YX, Jiang K, Wang ZQ, et al. Electronic structure and superconductivity in bilayer $La_3Ni_2O_7$. Phys Rev B 2024; 110: 205122.
[57] Julien MH, Campana A, Rigamonti A, et al. Glassy spin freezing and NMR wipeout effect in the high-Tc superconductor $La_{1.90}Sr_{0.10}CuO_4$: Critical discussion of the role of stripes. Phys Rev B 2001; 63: 144508.
[58] Chou FC, Borsa F, Cho JH, et al. Magnetic phase diagram of lightly doped $La_{2-x}Sr_xCuO_4$ from $^{139}$La nuclear quadrupole resonance. Phys Rev Lett 1993; 71: 2323.
[59] Li J, Ma P, Zhang H, et al. Pressure-driven right-triangle shape superconductivity in bilayer nickelate $La_3Ni_2O_7$. arXiv; 2404.11369; 2024.
[60] Wu G, Neumeier JJ, Hundley MF. Magnetic susceptibility, heat capacity, and pressure dependence of the electrical resistivity of $La_3Ni_2O_7$ and $La_4Ni_3O_{10}$. Phys Rev B 2001; 63: 245120.
[61] Hu Y, Zhang T, Zhao D, et al. Real-space observation of incommensurate spin density wave and coexisting charge density wave on Cr (001) surface. Nat Commun 2022; 13: 445.
[62] Dai P. Antiferromagnetic order and spin dynamics in iron-based superconductors. Rev Mod Phys 2015; 87: 855-896.
[63] Takabayashi Y, Ganin AY, Jeglič P, et al. The Disorder-free non-BCS superconductor $Cs_3C_{60}$ emerges from an antiferromagnetic insulator parent state. Science 2009; 323: 1585-1590.
[64] Khasanov R, Hicken TJ, Gawryluk DJ, et al. Pressure-induced split of the density wave transitions in $La_3Ni_2O_{7-\delta}$. arXiv: 2402.10485, 2024.

# Figures

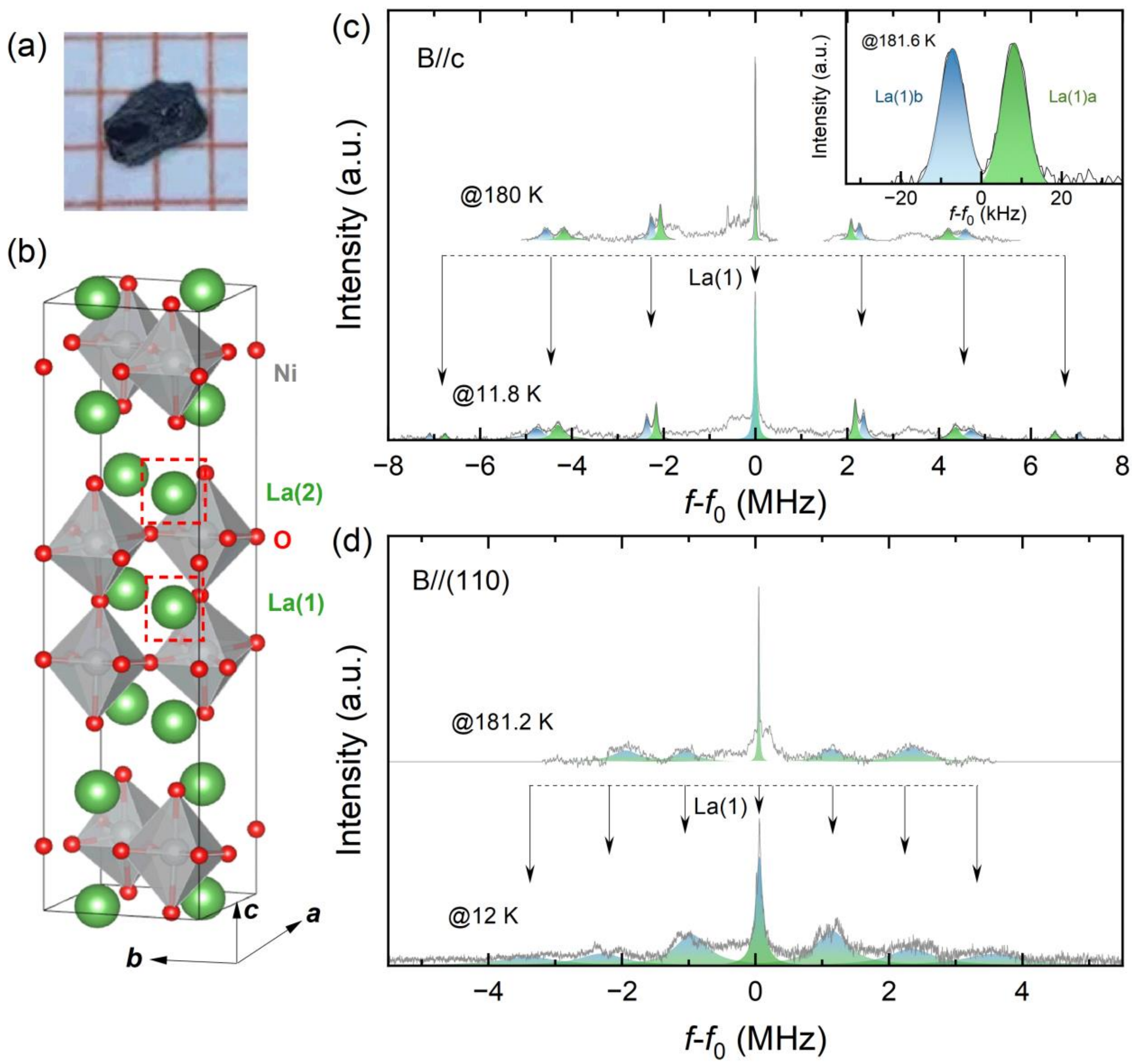

**FIG. 1** (a) The image of $La_3Ni_2O_7$ single crystal used for the present NMR study; (b) Sketch of the crystal structure of $La_3Ni_2O_7$; (c) and (d) Full NMR spectra of $^{139}$La nuclei above and below $T_{SDW}$ with external fields (c) $B \parallel c$ and (d) $B \parallel (110)$. The inset in (c) shows the central line of $^{139}$La above the SDW transition temperature.

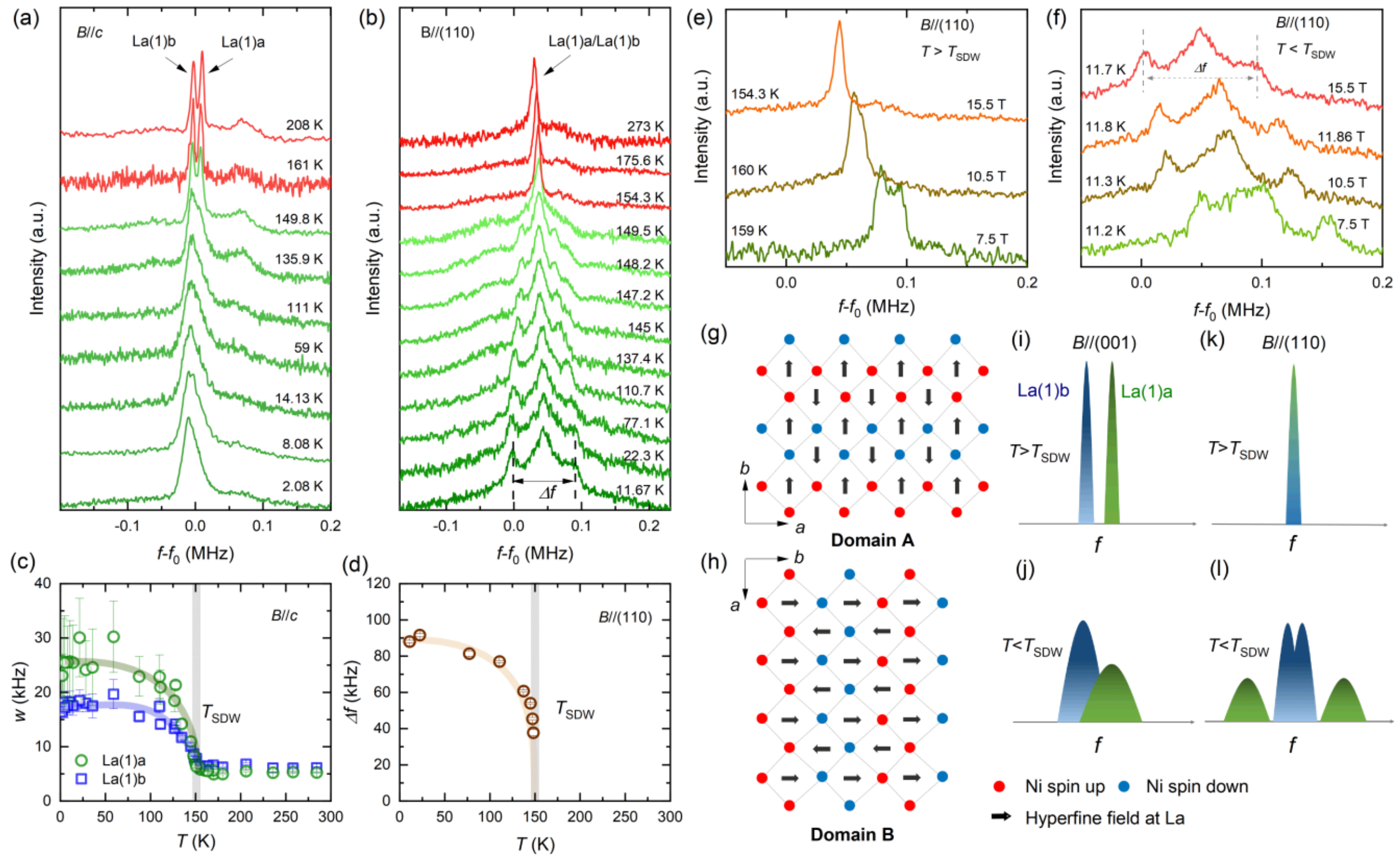

**FIG. 2** Temperature dependence of the $^{139}$La central transition lines for the (a) $B \parallel c$ axis and (b) $B \parallel (110)$ direction. (c) Temperature-dependent linewidth for the La(1)a and La(1)b sites. (d) Temperature-dependent magnetic splitting ($\Delta f$) of the La(1)b site for $B \parallel (110)$. The bold gray lines in (c) and (d) represent the SDW transition temperature at ~ 150 K. (e) and (f) Field dependence of the central line of $^{139}$La(1) with $B \parallel (110)$ (e) above and (f) below $T_{\text{SDW}}$. (g, h) Illustration of the relative orientations of the two twin domains, the double spin stripe model, as well as the transferred hyperfine fields at the La(1) sites. (i–l) The evolution of the simulated NMR spectra before and after the SDW transition. When the external field ($B$) is parallel to the $c$-axis, the hyperfine field $h_f$ at the La(1) sites is perpendicular to the external field $h_0$. The effective external field at the La(1) sites is expressed as $h_{\text{eff}} = \sqrt{h_{\text{f}}^2 + h_0^2}$, where $h_0$ is the applied external field and $h_{\text{f}}$ is the transferred hyperfine field due to magnetic order. Only spectral line broadening can be observed in this case. In contrast, when $B \parallel (110)$, the effective external field at the La(1) sites is expressed as $h_{\text{eff}} = h_0 \pm h_{\text{f}}$. A clear spectral splitting can be observed in this case.

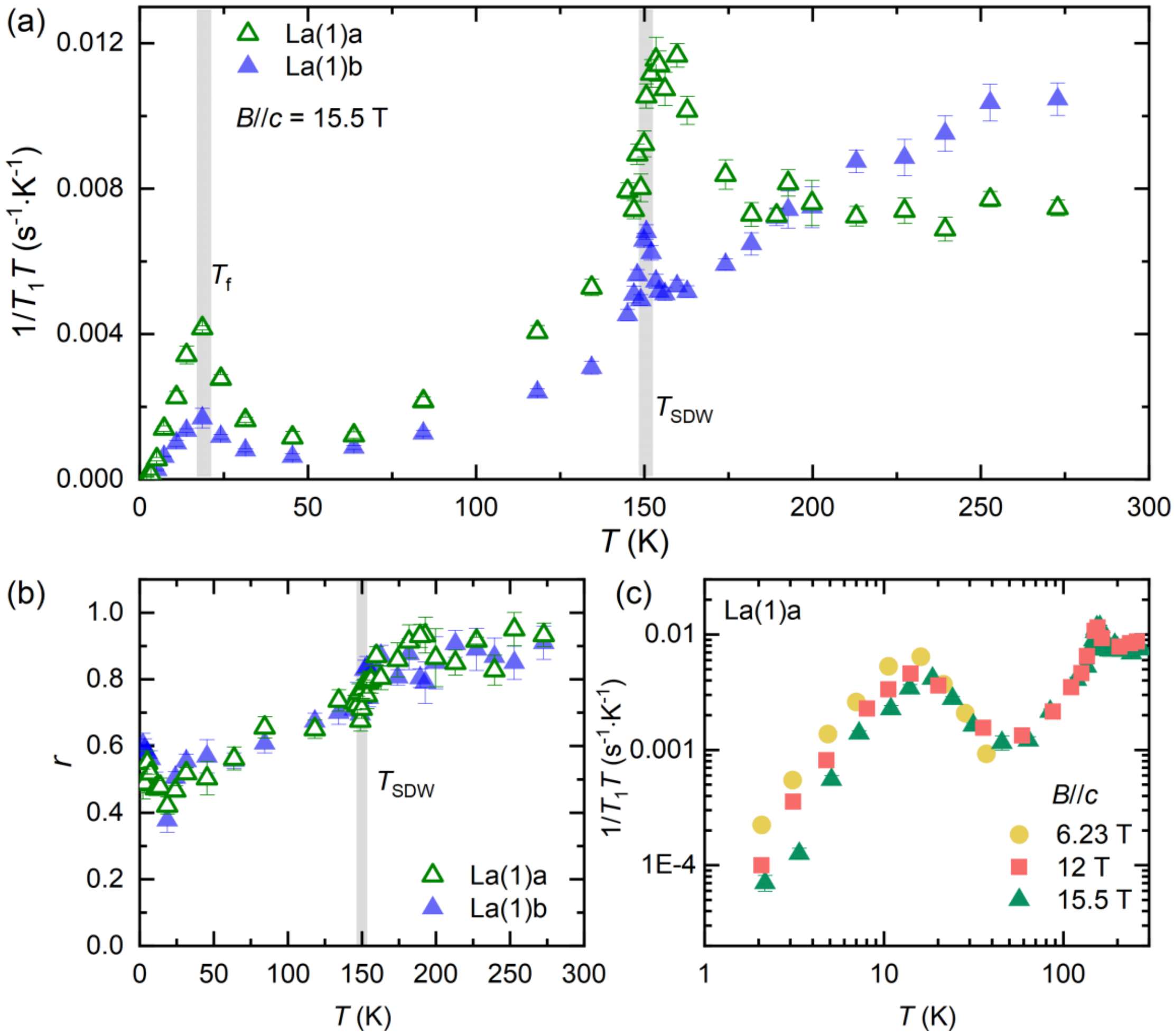

**FIG. 3** (a) Temperature dependences of $1/T_1T$ for both the La(1)a and La(1)b sites with $B \parallel c$ axis. The value of $T_1$ is extracted by fitting with the following formula $m(t) = m_0 + m_1[\frac{1}{84}e^{-\left(\frac{t}{T_1}\right)^r} + \frac{3}{44}e^{-\left(\frac{6t}{T_1}\right)^r} + \frac{75}{364}e^{-\left(\frac{15t}{T_1}\right)^r} + \frac{1225}{1716}e^{-\left(\frac{28t}{T_1}\right)^r}]$, where $r$ is the stretch exponent. (b) The temperature-dependent stretch exponent $r$ of $^{139}$La. (c) Field and temperature dependence of $1/T_1T$ for La(1)a.

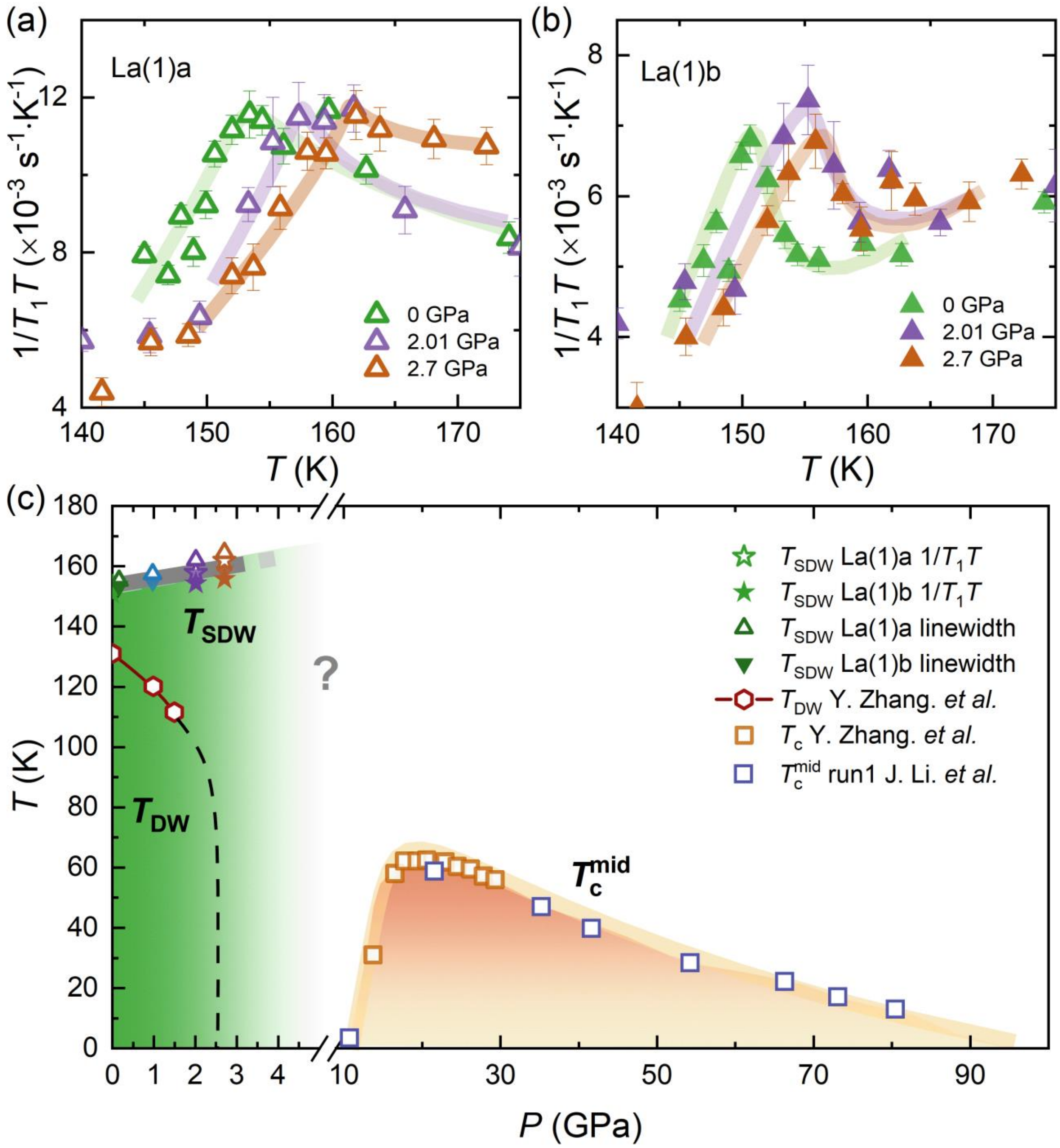

**FIG. 4** (a) and (b) Pressure-dependent $1/T_1T$ for both (a) La(1)a and (b) La(1)b sites with $B \parallel c$; (c) Pressure-dependent electronic phase diagram. The data for $T_{DW}$ and $T_c$ is taken from Refs. [41, 59], in which the measured single crystal of $La_3Ni_2O_7$ come from the same batch as the present work. Both of $T_{DW}$ and $T_c$ are determined by transport measurement. Here, the $T_{SDW}$ is determined by NMR measurement. Obviously, $T_{SDW}$ and $T_{DW}$ show different pressure-dependent behavior.